\documentclass[%
 reprint,
superscriptaddress,
 amsmath,amssymb,
 aps,
 prx,
]{revtex4-2}

\usepackage[english]{babel}
\usepackage{graphicx}
\usepackage{dcolumn}
\usepackage{bm}

\begin{document}


\title{Embedding Orthogonal Memories in a Colloidal Gel through Oscillatory Shear}

\author{Eric M. Schwen}
 \email{ems445@cornell.edu}
\affiliation{Department of Physics, Cornell University, Ithaca, NY 14850}
\author{Meera Ramaswamy}
\affiliation{Department of Physics, Cornell University, Ithaca, NY 14850}
\author{Chieh-Min Cheng}
\affiliation{Xerox Corporation, Rochester, NY 14605}
\author{Linda Jan}
\affiliation{Xerox Corporation, Rochester, NY 14605}
\author{Itai Cohen}
\affiliation{Department of Physics, Cornell University, Ithaca, NY 14850}

\date{\today}

\begin{abstract}
It has recently been shown that in a broad class of disordered systems oscillatory shear training can embed memories of specific shear protocols in relevant physical parameters such as the yield strain. These shear protocols can be used to change the physical properties of the system and memories of the protocol can later be “read” out. Here we investigate shear training memories in colloidal gels, which include an attractive interaction and network structure, and discover that such systems can support memories both along and orthogonal to the training flow direction. We use oscillatory shear protocols to set and read out the yield strain memories and confocal microscopy to analyze the rearranging gel structure throughout the shear training. We find that the gel bonds remain largely isotropic in the shear-vorticity plane throughout the training process suggesting that structures formed to support shear along the training shear plane are also able to support shear along the orthogonal plane. Orthogonal memory extends the usefulness of shear memories to more applications and should apply to many other disordered systems as well.

\end{abstract}

\maketitle

\section{\label{intro}INTRODUCTION}

Far from equilibrium, disordered systems can maintain memories of their preparation that can be “read out” using specific protocols \cite{keim_generic_2011, fiocco_encoding_2014}. When a memory is stored in a relevant physical parameter such as the yield strain, this memory encoding can be a useful tool for modifying the system properties. Oscillatory shear is an attractive method for embedding such memories and was first used for dilute systems of non-Brownian hard spheres \cite{corte_random_2008, pine_chaos_2005, nagasawa_classification_2019}. The logic in this case is relatively clear: shearing causes the particles to bump into each other and move apart until they find a reversible state where no collisions occur during a shear cycle. Further studies have argued that this oscillatory shear memory effect should apply to disordered systems in general—shearing causes rearrangement which continues until a reversible state is found \cite{paulsen_multiple_2014, sethna_deformation_2017, keim_multiple_2013}. Indeed, similar memory effects have been found in experiments and simulations of model amorphous systems \cite{fiocco_encoding_2014, adhikari_memory_2018, farhadi_shear-induced_2017}, granular systems \cite{royer_precisely_2015}, and glasses\cite{fiocco_memory_2015, parmar_strain_2019, hima_nagamanasa_experimental_2014, mukherji_strength_2019}. The interactions between particles can vary and even the nature of the “reversibility” can vary for different systems \cite{adhikari_memory_2018, regev_noise_2019, paulsen_minimal_2019, mungan_structure_2019, keim_global_2018, keim_memory_2019}. The core idea of particles rearranging and exploring possible states to find a reversible one still applies, regardless of the specifics of the system. 

Colloidal gels are another interesting candidate system for memory formation since they are amorphous and have a deformable network structure that can rearrange under applied strains. Previous studies have demonstrated use of both steady and oscillatory shear for tuning the properties of a colloidal gel, finding changes in the gel structure as well as the elastic modulus \cite{koumakis_tuning_2015, moghimi_colloidal_2017, rajaram_microstructural_2010, hsiao_role_2012, jamali_microstructural_2017, eberle_dynamical_2012, gordon_rheology_2016, conrad_structure_2008}. Here, we build on these results by investigating the regime of memory formation under low-amplitude oscillatory shear to determine whether the signal of a specific applied strain can be embedded and “read out” from a gel.

\section{\label{materials}Materials and methods}

For this experiment we chose a weak colloidal gel that allowed particle rearrangement under moderate shear. The colloidal gel consisted of monodisperse silica microspheres (diameter = 1.85 $\mu$m $\pm$ 0.08 $\mu$m) suspended in an index-matched solution of glycerol and water (75-25 by mass fraction). A small amount of fluorescein dye (2 mg/ml) was added to the solvent to allow imaging of the particles. Attraction was induced by the addition of magnesium sulfate salt (1M). The competition between van der Waals attraction and electrostatic repulsion creates a weak attraction between particles \cite{w._b._russel_colloidal_1989, eom_roughness_2017}. We chose particle volume fractions of $\phi \approx$ 35\% and 41\% which are sufficiently high to prevent the gel from collapsing under gravity.

This gel was sheared using a custom-built biaxial shear cell with a parallel plate geometry \cite{lin_multi-axis_2014, ramaswamy_how_2017, cheng_imaging_2011}. The shear cell was mounted on the stage of an inverted microscope (Zeiss Axiovert 200M) with a line-scanning confocal microscopy module (Zeiss LSM 5 Live). The top plate of the shear cell is a 3 mm x 3 mm silica wafer and the bottom plate is a glass coverslip. The gap between the top and bottom plates of the shear cell was set to 30 $\mu$m. This setup allows for rapid acquisition of three-dimensional stacks of images before, during, and after applied shear. The biaxial piezoelectric controller allows investigation of nontrivial shear patterns and orthogonal readout inaccessible by standard rotational rheometry.
The colloidal gel samples were vortexed for 2 minutes and sonicated for 30 minutes before loading into the shear cell. Before each trial, the gels were subjected to a sinusoidal shear rejuvenation procedure of 200\% strain applied at 20 Hz for 10000 cycles. The gels were then trained through the application of a sinusoidal strain pattern with a frequency of either 0.33 Hz or 0.50 Hz and a uniaxial strain amplitude $\gamma_0$ chosen for the specific trial. This oscillatory strain was repeated for 500-1000 cycles to allow the gels to reach steady state. Image stacks of the centers of the gels (64x64x6 $\mu$m) were taken stroboscopically (once per shear cycle) during training to analyze gel rearrangement (Fig. 1a). 
Stroboscopic image subtraction is our primary tool to quantify the amount of particle rearrangement between subsequent shear cycles (Fig. 1b). High levels of rearrangement show large differences between subsequent stroboscopic images, while subsequent images for a reversible state are nearly identical.  Therefore, we use the average magnitude of the image difference  $\langle | \Delta I | \rangle$ to quantify particle rearrangement over a single shear cycle.

\begin{figure}[t]
\includegraphics{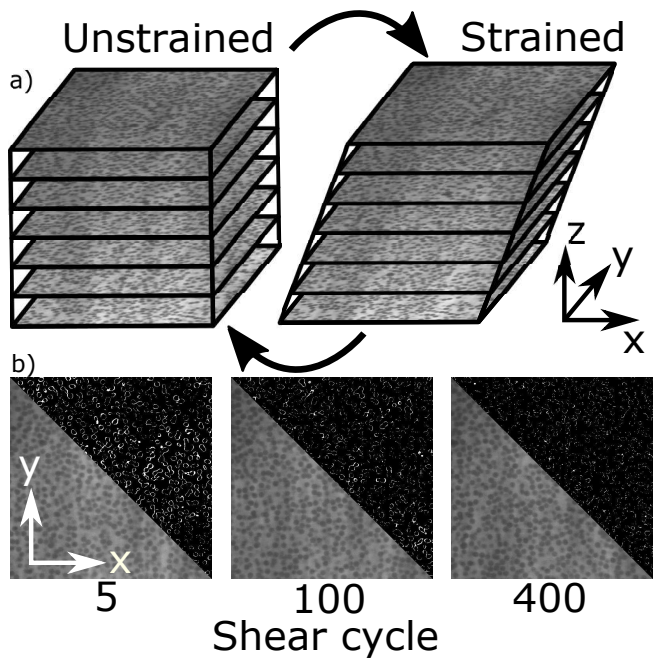}
\caption{\label{fig1}Gel training procedure. (a) An oscillatory strain pattern is applied to a colloidal gel while taking 3D confocal image stacks stroboscopically. Representative slices are shown from image stacks in both unstrained and strained positions. (b) Split images show confocal images in the lower left and image difference in the upper right. Snapshots show different shear cycles during gel training. To obtain image difference, each image is filtered to remove noise and identify regions where particles are present. The later image is then subtracted from the previous cycle’s image and the absolute value of this result is the image difference. High intensity areas in the image difference identify regions where a particle is present in one image but not in the other.}
\end{figure}

We used a strain amplitude sweep to compare the gel response before and after training and assess the effect of the training procedure. For each strain sweep a selection of $\sim$ 15 strain amplitudes $\gamma$ ranging from well below to well above the training strain $\gamma_0$ were applied to the gel. Beginning from the lowest strain and proceeding linearly, five cycles at each strain amplitude were applied while imaging stroboscopically. Analyzing these images allows a measurement of the gel response at different strain amplitudes below and above the training strain.

\section{\label{results}Results}
A core parameter controlling memory formation in disordered systems is the training strain amplitude $\gamma_0$.  We trained a colloidal gel using a variety of different strain amplitudes and observed significant differences in both particle rearrangement during training and in memory readout from strain amplitude sweeps (Fig 2). 

For low strains ($\gamma_0 \approx$ 0.06) the gel does not need to restructure extensively from its initial configuration and quickly reaches a reversible state. Figure 2a shows an example of the image difference $\langle | \Delta I | \rangle$ over time where the system reaches the noise floor of our measurements early in the training process. Strain amplitude sweeps (Fig. 2b) show significantly less rearrangement in the trained gel. While the effect of training at such small amplitudes is modest, the changes observed are significantly different from those for a gel that simply ages without shear for the same period of time (shown in Supplemental Material \cite{supp}).

Moderate strains ($\gamma_0 \approx$ 0.19) cause much more rearrangement and show clearer signals of both gel training and the formation of a memory of a specific training strain. The amount of rearrangement decreases with additional training cycles and approaches a reversible state with measured image difference at the noise floor (Fig. 2c). The strain amplitude sweep for the trained gel shows little to no rearrangement below the training strain—the gel can support these small strains without rearrangement (Fig. 2d). For strains larger than the training strain $\gamma_0$, the gel begins to rearrange again. The training procedure is effectively able to set a new, specific yield strain for the gel at the training strain. The training strain can be read out from the bend in the image difference data where the image difference rises above the noise floor in either Figure 2b or Figure 2d. This key result clearly demonstrates that our gel can form a memory of the training strain that can be read out using a strain amplitude sweep.

For higher training strains ($\gamma_0 \approx$ 0.21) the gel rearrangement decreases over time but approaches a steady state image difference well above the noise floor (Fig. 2e). The gel rearranges indefinitely because the applied strain is too large to reach a reversible state. The strain amplitude sweep does show less rearrangement for the trained gel at every strain amplitude. However, the gel yields and rearranges even at low strains and the applied training strain cannot be easily read from the results (Fig. 2f). For even higher training strains, no evidence of the training is measurable. The image difference does not change throughout training and the strain amplitude sweeps show no difference between the trained and untrained gels (shown in Supplemental Material \cite{supp}).

\begin{figure*}[t]
\includegraphics{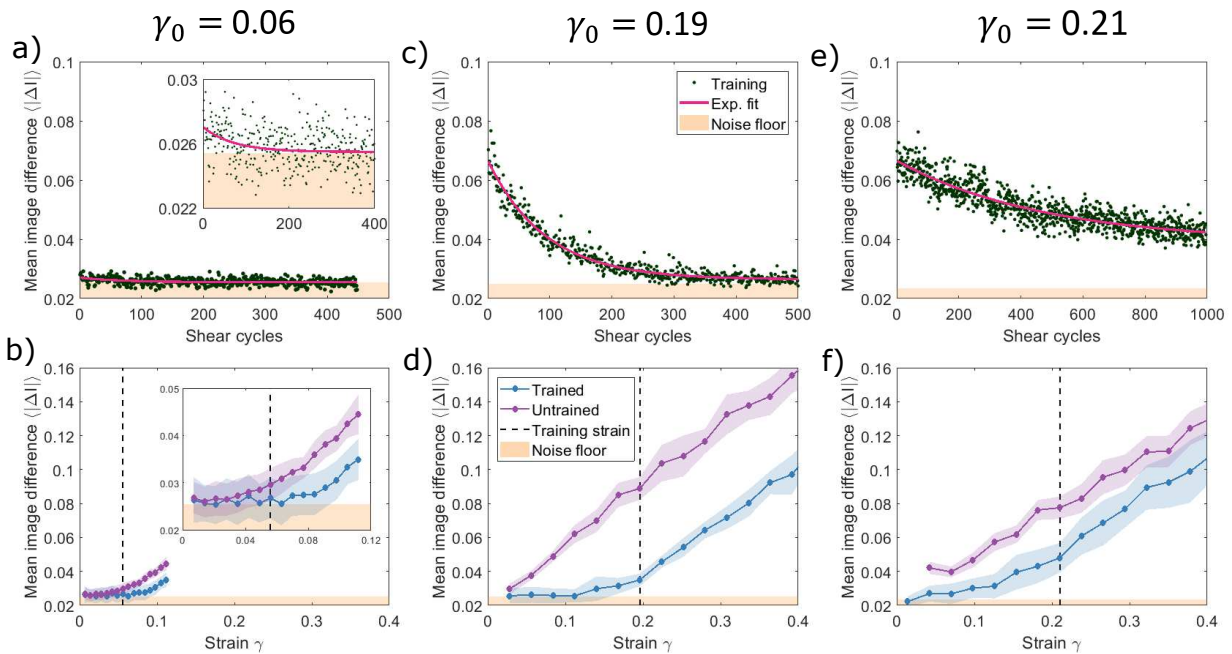}
\caption{\label{fig2}Gel training results. Top row: Average magnitude of the mean image difference $\langle | \Delta I | \rangle$ during the training process as a function of shear cycles for training strain $\gamma_0$ and frequency (a) $\gamma_0 = 0.06$ at 0.5 Hz, (c) $\gamma_0= 0.19$ at 0.33 Hz, and (e) $\gamma_0 = 0.21$ at 0.33 Hz. Pink lines show an exponential fits of the form $\langle | \Delta I | \rangle = a\exp(-t/\tau)+b$ where $t$ is the number of shear cycles, $\tau$ is the characteristic training time, and $a$ and $b$ are constants. Bottom row: Strain amplitude sweeps for the gels above comparing the response of the gel to different strains before and after training. Insets show the same data with constrained axes for clarity. Results are shown for a $\phi = 35\%$ gel. Similar results are obtained for $\phi=41\%$.}
\end{figure*}

These data suggest there exists a critical training strain $\gamma_0^c$ below which the gel can reach a reversible state and above which it rearranges indefinitely. Recent work [3] in memory-forming disordered systems has found power law scaling for the characteristic number of shear cycles $\tau$ required to reach a steady state following a form $\tau \sim | \gamma_0 - \gamma_0^c | ^{-\nu}$ where $\gamma_0^c$ is the critical strain representing the maximum effective training strain and the exponent $\nu \sim 1.3$. We do find support for this idea in our gel measurements over tens of trials using volume fractions of 35\% and 41\% as well as oscillatory shear frequencies of 0.33 Hz and 0.50 Hz (shown in Supplemental Material \cite{supp}). However, experiments aimed at measuring the precise value of the critical strain, require long times of $\sim$ 8 hours over which we observed drift in the particle properties and the critical strain $\gamma_0^c$. We also found that $\gamma_0^c$ varies between preparations. Finally, while we only report measurements where wall slip was constrained below 20\% throughout the training and readout processes, this small slip can also affect measurement of the divergence at $\gamma_0^c$. Therefore experiments with even more stable systems or simulations will be needed to determine the exact value of the critical strain and power law exponent. 

The different shear response between the trained and untrained gel suggests a change in the gel microstructure \cite{hexner_linking_2018, hexner_role_2018, tsurusawa_direct_2019,zia_micro-mechanical_2014, rajaram_dynamics_2011, patinet_connecting_2016}. We use confocal microscopy to image the gel structure and examine how the trained gel is able to support larger strains without rearranging. We take 3D image stacks (64x64x30 $\mu$m) before and after training as well as during pauses between shear cycles during the training process. We then locate particles using both a centroid-based method \cite{d._b._allan_trackpy_2014} and parameter extraction from reconstructing images \cite{bierbaum_light_2017} (PERI). Centroid-based methods can rapidly locate many particles in a large image volume and provide good statistics for structure measurements averaged over the system. PERI enables measurement of particle positions and radii on the nm scale and allows the precise local structure measurements. Extracting the precise particle positions and sizes allow us to use surface separation to measure nearest neighbor distributions and determine their influence on which particles are likely to rearrange.
Often, the pair distribution function provides a first measure of large structural changes in a gel. Indeed, previous studies of colloidal gels under shear have noted substantial densification of the gel \cite{koumakis_tuning_2015, moghimi_colloidal_2017, rajaram_microstructural_2010, chan_two-step_2012} as a significant or even dominant factor in the rheological response of the gel. In these studies, large dense clusters and corresponding voids developed as the gel was sheared. Here, however, the lack of substantial densification is immediately apparent through visual inspection of the gel images before and after training—they are nearly indistinguishable (Fig. 1). As such, the structural modifications are significantly smaller and the pair distribution function only shows a slight increase in the primary peak (Fig. 3a). 

\begin{figure*}[t]
\includegraphics{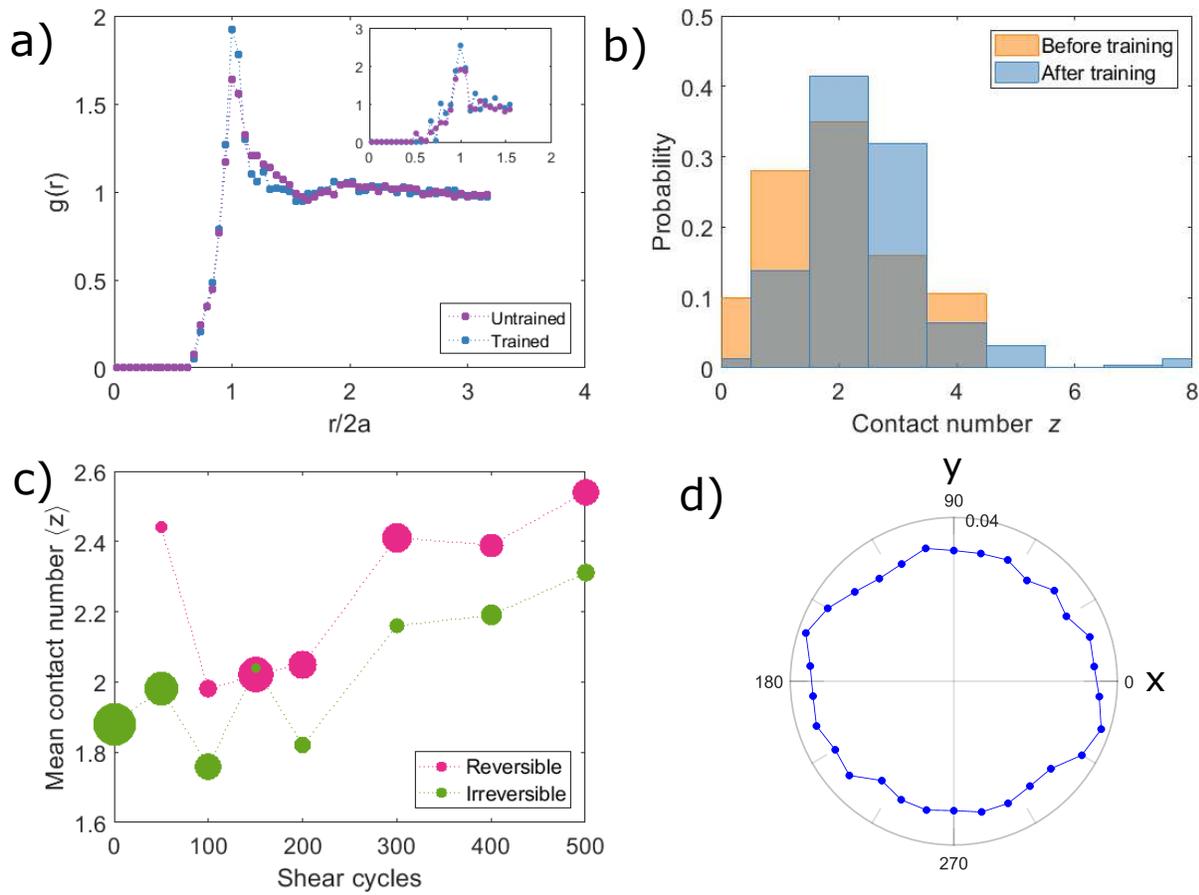}
\caption{\label{fig3}Gel structure analysis. (a) Pair distribution function for a representative gel sample before and after training. Main plot uses positions from centroid-based tracking methods. Inset uses positions from PERI which is more precise but tracks fewer particles. (b) Contact number distributions for trained and untrained gel. Contacts are defined by a particle surface separation less than 50 nm as determined using PERI. (c) Mean contact number $\langle$z$\rangle$ plotted as a function of shear cycle for both reversible and irreversible particles. Dot size represents number of particles in the classification. (d) Bond angle probability distribution in the shear-vorticity (x-y) plane for a trained gel. Bond angles for this plot are determined for particles with center-to-center separation less than 2.1 $\mu$m. Similar results are found when bond angles are determined by surface separation (shown in Supplemental Material \cite{supp}). Results are shown for a $\phi= 35\%$ gel for training strain $\gamma_0=0.06$  at 0.33 Hz. Similar results are obtained at other training conditions.}
\end{figure*}

This change in the primary peak indicates the contact number z is changing. Such contacts are important for the mechanical stability within gels and have been correlated with both gel aging \cite{zia_micro-mechanical_2014} and the gel’s ability to sustain larger strains \cite{hsiao_role_2012}. We define particles to be in contact if their surface-to-surface separation is less than 50 nm. Using a different separation cutoff does slightly shift the mean contact number (by $\sim$ 0.5) but does not change the observed trends. We plot the probability distributions for contact number z for a gel before and after training in Figure 3b. After training, the gel shows a shift towards higher contact numbers. 

We further analyze the data to determine how the contact number is related to particle rearrangements during training \cite{van_doorn_linking_2017, dibble_structural_2008}. It is not a priori clear how these two quantities should be related. On the one hand a higher contact number indicates a more constrained environment that is less able to strain without rupture. On the other hand a network that has too low of a contact number may not be stable enough to thermal fluctuations once strained. To investigate the relationship between contact number and reversibility, we locate the particles and their neighbors as before, but now track their positions between stroboscopic images to obtain a net displacement for each particle. Using the noise floor established from an unsheared gel as a cutoff, we label any particle displacing more than 60 nm as an irreversible particle. The contact number distributions for the two populations show a significant difference: irreversible particles tend to have fewer contacts. Put the other way, particles with fewer contacts are more likely to rearrange in a shear cycle. Plotting the mean contact number $\langle$z$\rangle$ at different training cycles shows that this trend of irreversible particles having fewer contacts is generally true throughout the training process (Fig. 3c). Additionally, the population of irreversible particles decreases as training proceeds and the gel approaches a reversible state. Similar results are found for the Voronoi volumes of reversible and irreversible particles (shown in Supplemental Material \cite{supp}). These observations suggest a picture where irreversible particles with fewer contacts rearrange until they find a stable configuration with more contacts and join the reversible population. The end result is a trained gel with a higher contact distribution as seen in Figure 3a. 

Bond angles can also provide meaningful information about how a gel will respond to external shear. Colloidal gels under shear will sometimes arrange into chains aligned in the shear plane \cite{rajaram_microstructural_2010}. However, measurements of our colloidal gels both before and after training show a largely isotropic distribution of bond angles in the shear-vorticity plane (Fig. 3d). This lack of directionality suggests an interesting idea: if the structures formed in the trained gel also extend into the orthogonal direction, the gel may also be trained to support orthogonal shear. 

In order to test the idea of orthogonal memories, we train a gel as usual by applying uniaxial shear along the x-z shear plane. For the strain amplitude sweep, however, we alternate between applying the shear along the x-z shear plane and the orthogonal y-z plane. The comparison of mean image difference before and after training shows a remarkable result—the gel is also trained to support orthogonal strains (Fig 4). Even though the training flows were directed along the x-z plane, there is no significant difference in the trained gel response for strains applied along the x-z and y-z shear planes. This result, in combination with the even distribution of bond angles, suggests that the structures formed to support the training strain extend and support in three dimensions rather than being concentrated along the shear-gradient plane. The lack of a strong directionality also suggests that the memory effect may be robust and relevant for a variety of shear procedures that may arise in industrial use. 

\begin{figure}[t]
\includegraphics{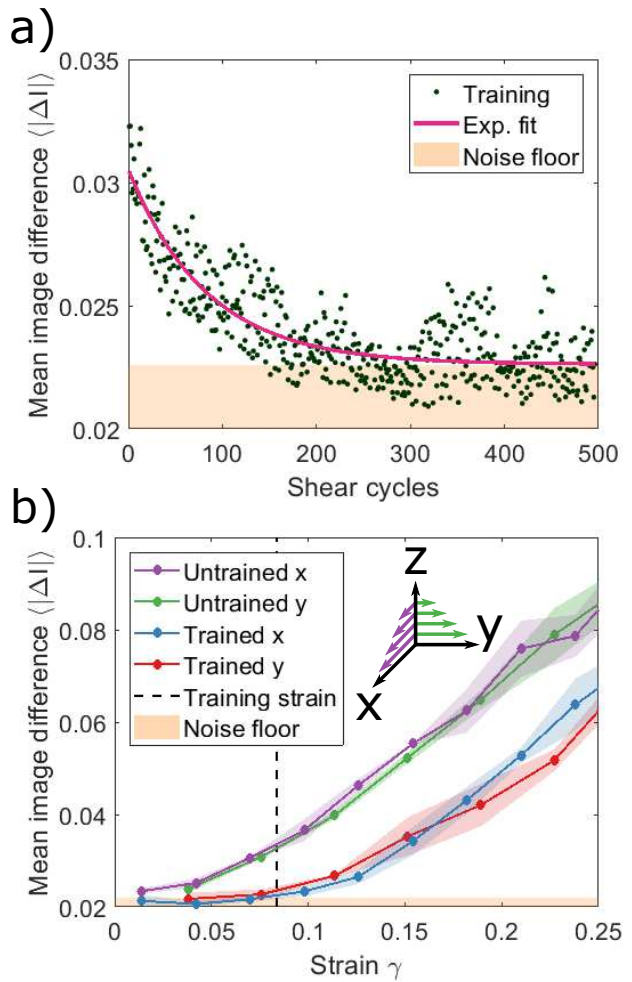}
\caption{\label{fig4}Orthogonal training. (a) Mean image difference $\langle | \Delta I | \rangle$ as a function of shear cycles during training for training strain $\gamma_0=0.08$ at 0.33 Hz. (b) Strain amplitude sweeps comparing the response of the gel to along the x-z and orthogonal y-z shear planes before and after training.}
\end{figure}

\section{\label{discussion}Discussion}

Gels are ubiquitous in many applications ranging from beauty aids to biological gels to printer toners. The properties of a gel, however, are usually determined by the constituent materials and cannot be easily adjusted. Our measurements show that oscillatory shear training can be used to embed a memory of a specific strain in a colloidal gel. This training provides a tool to control gel properties such as the gel connectivity and yield strain without changing the materials. 

Colloidal gels provide a compelling addition to the memory effect story because they include an attractive potential between particles and a network structure. Our precision measurements of particle positions and radii \cite{bierbaum_light_2017} enabled us to show that the transition to a reversible configuration in our gels coincides with an increase in nearest-neighbor contacts at the individual particle level. Further measurements of the response of the entire network throughout a shear cycle would help identify whether the increase in contact number stabilizes the gels by increasing strand thickness or enhancing crosslinking between strands \cite{c.johnson_yield_2018, colombo_stress_2014, van_doorn_strand_2018}. Other aspects of shear training memories could be addressed by adjusting the shearing frequency or gel volume fraction \cite{smith_yielding_2007}. Both parameters are likely to change the critical training strain and the characteristic time to reach steady state. Additional studies could also relate memory formation in gels to changes in the viscoelastic moduli. Previous works have linked shear-induced changes in microstructure to changing moduli \cite{koumakis_tuning_2015, moghimi_colloidal_2017}. Here, the fluid shear stresses dominate the elastic response of the weak gel so moduli changes are not easily measurable. Such changes in moduli could, however, become more dominant in denser gels.

More broadly, it would be interesting to determine whether our findings for colloidal systems extend to other systems. For example, it is possible that biological systems use oscillatory mechanisms to embed physical memories in the gels that give rise to the mechanical properties of cells and the extracellular matrix in which they reside. Rheological properties of biopolymer gels such as strain-stiffening and negative normal stresses have been linked to functions such as resisting strain injury and facilitating organelle motion \cite{janmey_negative_2007, storm_nonlinear_2005}. Repetitive oscillations or shear training in biological gels could be similarly linked to facilitating operations under shear strains.

Finally, the discovery that gels can support memories in multiple shear planes despite training along a single flow direction indicates that coupling in systems that display memory may be tunable through the degree of bond isotropy. In our gel where the bonds are evenly distributed in the shear-vorticity plane, we find nearly perfect coupling of memories along different shear planes. Systems with similar isotropy, such as non-Brownian hard spheres, are expected to show similar orthogonal memory effects. In systems where such symmetries are broken, however, memories along orthogonal planes may be reduced or even eliminated providing an additional handle for tailoring the material response. Lastly, it would be interesting to consider the effect of system isotropy on memory formation in more abstract contexts. For example, in machine learning processes there are techniques for transfer learning where training a model on one task allows for solving problems in a separate but related task \cite{bengio_representation_2013}. As we learn more about memory formation in different contexts and understand its fundamental underpinnings, this approach may emerge as a powerful method for modifying and even controlling the properties of systems both physical and abstract.

\begin{acknowledgments}
The authors would like to thank the Cohen lab group members for helpful discussions. This research was supported in part by the National Science Foundation under NSF CBET-PMP Award No. 1509308 and Grant No. NSF PHY17-48958. This research was also supported in part through funding from Xerox Corporation to the Cornell Center for Materials Research.
\end{acknowledgments}


%

\end{document}



\title{Supplemental Information for ``Embedding Orthogonal Memories in a Colloidal Gel through Oscillatory Shear"}

\author{Eric M. Schwen}
 \email{ems445@cornell.edu}
\affiliation{Department of Physics, Cornell University, Ithaca, NY 14850}
\author{Meera Ramaswamy}
\affiliation{Department of Physics, Cornell University, Ithaca, NY 14850}
\author{Chieh-Min Cheng}
\affiliation{Xerox Corporation, Rochester, NY 14605}
\author{Linda Jan}
\affiliation{Xerox Corporation, Rochester, NY 14605}
\author{Itai Cohen}
\affiliation{Department of Physics, Cornell University, Ithaca, NY 14850}

\date{\today}

\maketitle

\beginsupplement

\section{\label{training}Training strain extremes}

\begin{figure*}[b]
\includegraphics{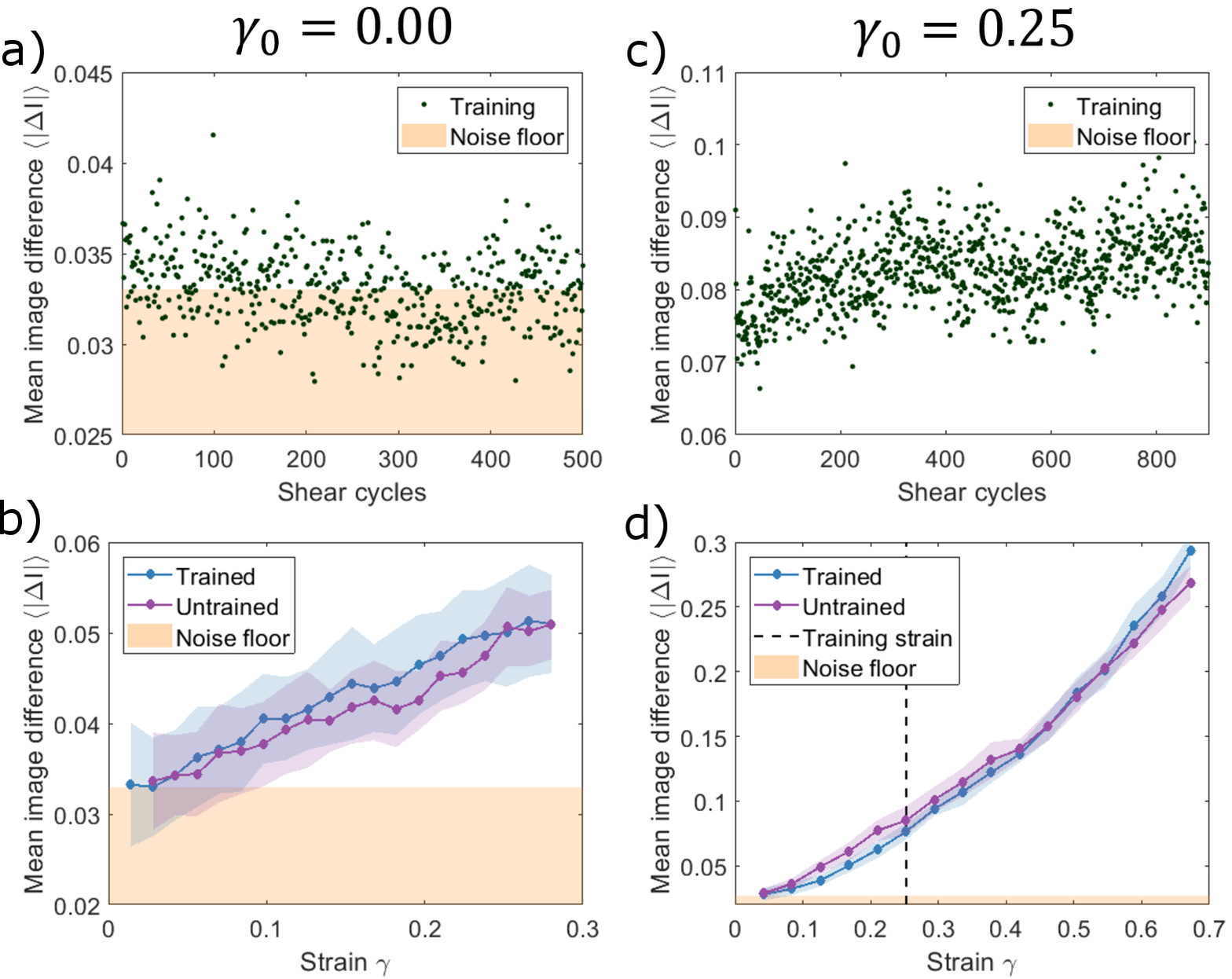}
\caption{\label{si_fig1} Mean image difference $\langle | \Delta I | \rangle$ as a function of time while waiting without shearing. X-axis shows equivalent time for “shear cycles” at 0.2 Hz following the form of similar plots. (b) Strain amplitude sweep at 0.2 Hz comparing the shear response of the gel before and after waiting. (c) Mean image difference $\langle | \Delta I | \rangle$ as a function of shear cycles for training strain $\gamma_0=0.25$ at frequency 0.5 Hz. (d) Strain amplitude sweep comparing the shear response of the gel from (c) before and after training.}
\end{figure*}

Measurements of the gel response for zero training strain show that no memory is formed without shear training. To measure the gel response at $\gamma_0=0$, we let the gel sit without shearing for the same period of time as the normal shear training procedure. As it sits, we image the static gel at the same frequency we normally would for stroboscopic images during training. The image difference over time is plotted in Figure \ref{si_fig1}a and shows no significant change over time. Strain amplitude sweeps taken as normal before and after this wait period and show no significant difference (Fig. S1b). No “memory” is formed in the gel and no measurable changes are observed. This check confirms that our measured memory effect does not emerge through normal gel aging. 

Image difference for very high training strains show another expected result—no memory is formed if the training strain is too high for stable structures to form. The image difference over time for $\gamma_0=0.25$ at 0.5 Hz shows no net decrease over the training period (Fig. S1c). The gel continues the same amount of rearrangement as long as it is sheared. The strain amplitude sweep shows no significant difference between the untrained and trained gel (Fig. S1d). This null result is expected for high enough training strains. The large strain rips apart gel structures and prevents the formation of structures that could shear without rearranging.

\section{\label{characteristic}Characteristic training time}

\begin{figure*}[h]
\includegraphics{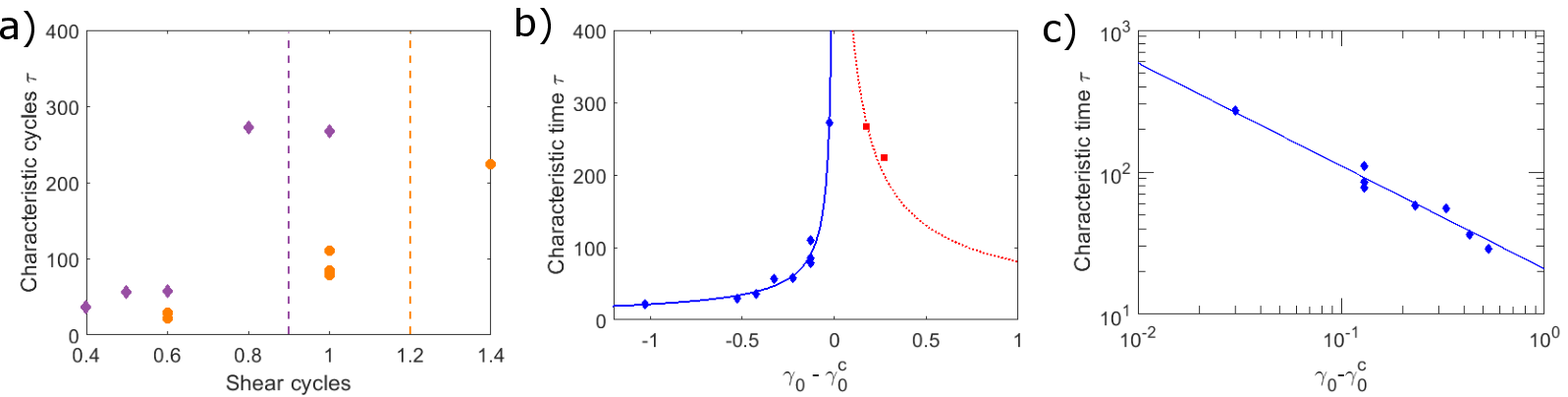}
\caption{\label{si_fig2}Characteristic training time. (a) Characteristic training time $\tau$ plotted vs. training strain $\gamma_0$ for different sample preparations at $\phi=0.35\%$. Dotted vertical lines mark the calculated critical strain $\gamma_0^c$ for each preparation. (b) Combined plot of characteristic training times $\tau$ for different sample preparations. Solid blue line shows the power-law fit $\tau \sim | \gamma_0 - \gamma_0^c | ^{-\nu}$ with $\nu=0.7 \pm 0.2$. Dotted red line sketches decreasing $\tau$ at high strains. (c) The same data for $\gamma_0 < \gamma_0^c$ plotted on a log-log scale}
\end{figure*}

For our gels, the mean image difference $\langle | \Delta I | \rangle$ during shear training decreases towards a steady-state value at long times. The decrease is well fitted by the form $\langle | \Delta I | \rangle = ( \langle | \Delta I | \rangle_0 - \langle | \Delta I | \rangle_\infty ) \exp(-t/\tau) + \langle | \Delta I | \rangle_\infty$ where $\langle | \Delta I | \rangle_0$ and $\langle | \Delta I | \rangle_\infty$ are the initial and steady-state image difference values, t is time measured in shear cycles, and $\tau$ is the characteristic time to reach steady state (see main Fig. 2). The characteristic time $\tau$ varies with training strain amplitude $\gamma_0$ and appears to diverge at a critical training strain $\gamma_0^c$. This critical training strain marks the division between reversible steady states  below and steady states with rearrangement above. The characteristic time divergence is consistent with power-law scaling of the form $\tau \sim | \gamma_0 - \gamma_0^c | ^{-\nu}$ similar to that seen in other memory-forming disordered systems. However, experiments to precisely measure the critical training strain $\gamma_0^c$ and scaling exponent $\nu$ require long times of $\sim$ 8 hours over which we observed drift in particle properties. This time constraint limited any set of measurements to a maximum of $\sim$ 6 different training strains. Additionally, the critical strain seemed to vary between preparations. In order to compare different preparations, we fit the characteristic times for each preparation to the scaling form $\tau \sim | \gamma_0 - \gamma_0^c | ^{-\nu}$ and determined the critical strain $\gamma_0^c$ (Fig. S2a). We then combined the data sets from the different preparations to measure the critical exponent $\nu=0.7 \pm 0.2$ (Fig. S2b). The limited time for trials and variation between preparations prevent easy measurement of a more precise critical exponent with this gel system.

\section{\label{voronoi} Voronoi volumes}

\begin{figure*}[h]
\includegraphics{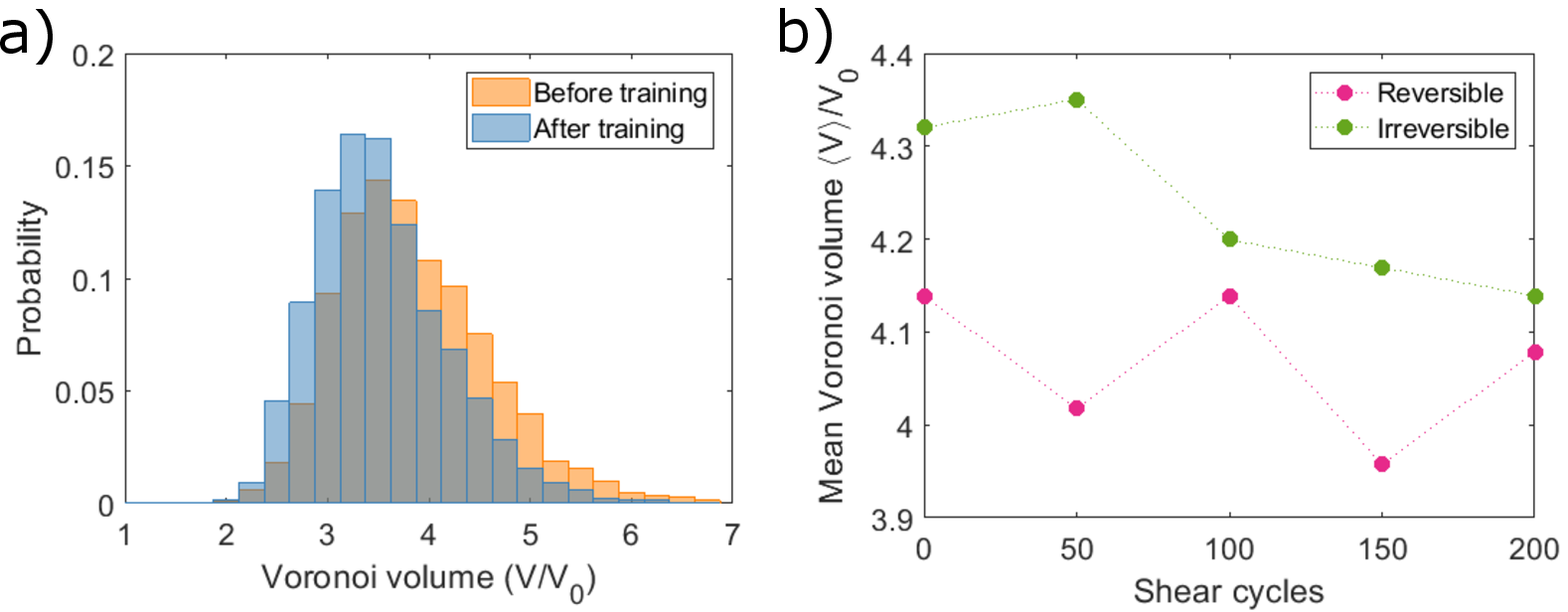}
\caption{\label{si_fig3}(a) Distribution of Voronoi volumes $V$ before and after training scaled by single particle volume $V_0$. (b) Mean Voronoi volume scaled by single particle volume $V_0$  for reversible and irreversible particles at different points in training process. Results are shown for a $\phi= 35\%$ gel at training strain $\gamma_0=0.06$  at 0.33 Hz with particles located using centroid-tracking methods.}
\end{figure*}

An alternative method of analyzing nearest neighbors and local structure relies on Voronoi tessellation. For Voronoi tessellation, the full 3D space of the image is divided into volumes based on the distance to the gel particles. Each particle has a Voronoi volume made up of all points closer to to it than to any other particle. The magnitude of that volume can serve as a measure of the amount of free space around the particle. 

We calculate the Voronoi volumes for particles in our gel before and after shear training as well as for pauses during the training process. After training, the distribution of Voronoi volumes peaks at a lower value with a longer tail extending to high volumes. This observation suggests slight densification where the gel divides into more dense regions and larger voids (Fig. S3a). 
Returning to the designation of reversible and irreversible particles can give some insight into how Voronoi volume relate to particle rearrangement. We find that irreversible particles tend to have a larger Voronoi volume than reversible ones (Fig. S3b). This pattern continues during the training process and fits in with our contact number observations. Particles with many close neighbors will have small Voronoi volumes and will not rearrange. The particles with larger Voronoi volumes are more likely to be on the edges of structures and are more likely to move between cycles. 

\section{\label{bond} Bond angles}

\begin{figure*}[h]
\includegraphics{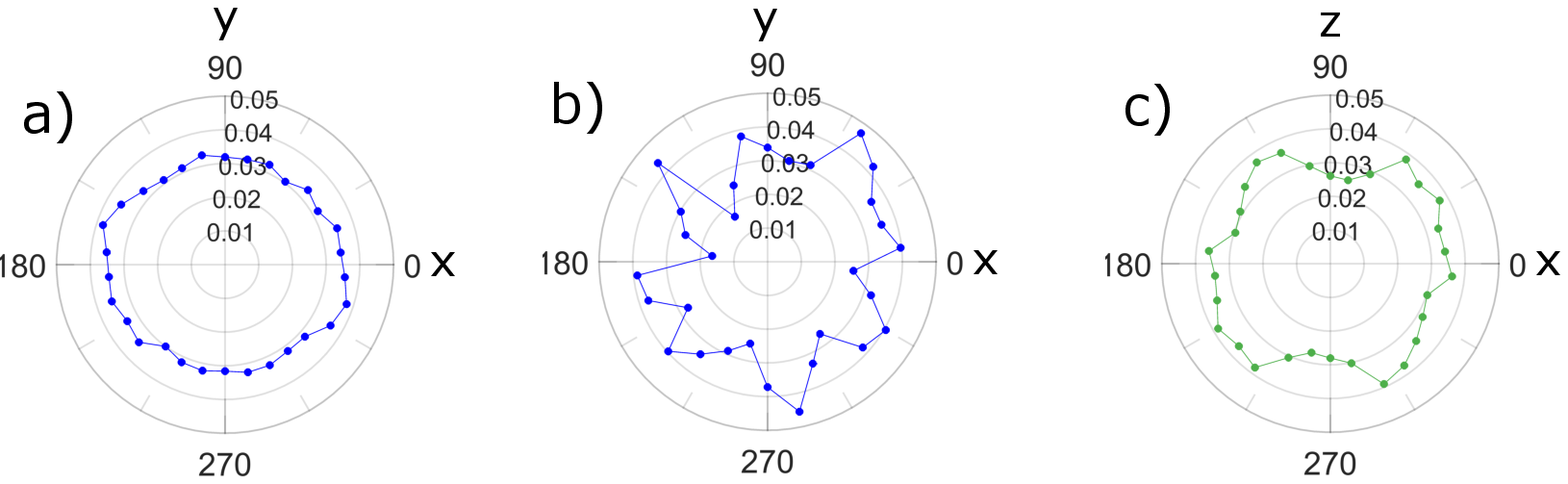}
\caption{\label{si_fig4}(a) Probability density function for gel bond angles after training projected into the shear-vorticity plane. Bond angles for this plot are determined for particles with center-to-center separation less than 2.1 $\mu$m. (b) Same probability density function as (a) but using a smaller subset of particles located precisely using PERI. Variation from isotropic distribution is within noise. (c) Probability density function for gel bond angles after training projected into the shear-gradient plane. Bond angles for this plot are determined for particles with center-to-center separation less than 2.1 $\mu$m. These data are for a gel after training with $\gamma_0=0.14$ at 0.33 Hz and are representative of the results we find for other parameters.}
\end{figure*}

The distribution of bond angles can show whether particles are preferentially forming chains in response to shear flow. We locate particles using centroid-based methods and then parameter extraction from reconstructing images (PERI). Using centroid-based methods, we locate all particles in 3D image stacks (64x64x6 $\mu$m) and define neighboring particles to be in contact if their center-to-center distance is less than 2.1 $\mu$m in order to account for the contact range but also allow for polydispersity and uncertainties in particle positions. A sample probability density function for bond angles of a trained gel projected into the shear-vorticity plane is shown in Figure S4a. A preference for particles chaining along the shear flow direction would appear as a peak at $0^\circ$. We instead see an isotropic distribution of bond angles, suggesting that the reversible structures formed under training extend into the shear-vorticity plane as well. Similar results are found for particles located using PERI where we define bonds as particles with surface separations less than 50 nm (Fig. S4b). Noise is much higher due to the smaller subset of particles located, but the lack of a preferential bond direction remains.

The bond angle distribution projected into the shear-gradient plane is plotted in Figure S4c and shows a non-isotropic structure. Particles tend not to stack directly on top of each other, as shown by the low probability of finding a bond at $0^\circ$ or $180^\circ$. A similar distribution shape is found both before and after training, suggesting that it is not the result of the shear training process. It may instead arise from gravitational effects or the relatively narrow gap between the top and bottom plates (15 particle diameters compared to hundreds in the other directions).